\documentclass[10pt,journal]{IEEEtran}

\usepackage{amsmath,amssymb,amsfonts,amsthm,bm}
\usepackage{graphicx}
\usepackage{subcaption}
\usepackage[labelsep=period]{caption}
\usepackage[dvipsnames,table]{xcolor}
\usepackage{cite}
\usepackage{booktabs}
\usepackage{siunitx}
\usepackage{stfloats}
\usepackage{balance}

\usepackage{algorithm}
\usepackage{algpseudocode}

\IEEEoverridecommandlockouts
\allowdisplaybreaks[4]

\newcommand{\Si}{\mathrm{Si}}
\newcommand{\Ci}{\mathrm{Ci}}

\def\BibTeX{{\rm B\kern-.05em{\sc i\kern-.025em b}\kern-.08em
		T\kern-.1667em\lower.7ex\hbox{E}\kern-.125emX}}

\begin{document}

\title{Rotatable Coupler Antenna Enhanced Wireless Network: Modeling and Coupler Rotation Optimization}



	
	
\author{{Xiaodan Shao, \IEEEmembership{Member,~IEEE}, Chuangye Shan, Weihua Zhuang, \IEEEmembership{Fellow, IEEE}, Xuemin (Sherman) Shen, \IEEEmembership{Fellow, IEEE}}
	\thanks{X. Shao, W. Zhuang, and X. Shen are with the Department of Electrical and Computer Engineering, University of Waterloo, Waterloo, ON N2L 3G1, Canada (E-mail: \{x6shao, wzhuang, sshen\}@uwaterloo.ca)}
	}
\maketitle
	
	\begin{abstract}
Flexible coupler antenna systems have recently received significant research interest due to their capability to intelligently reconfigure wireless channels by controlling coupler positions and/or rotations and dynamically exploiting mutual coupling. In this paper, we investigate a new type of flexible coupler antenna, termed rotatable coupler antenna (RCA), for enabling spectrum and energy efficient wireless communication cost-effectively. 
Specifically, an RCA consists of one fixed active antenna and multiple low-cost passive couplers, each of which can independently rotate in three-dimensional (3D) space, so as to collaboratively achieve mechanical beamforming without requiring additional radio-frequency (RF) chains for the couplers. We study an RCA-enhanced point-to-point communication system, where one RCA is deployed at the transmitter to serve a single user equipped with a fixed antenna. Based on multi-port circuit theory, we establish the channel model and characterize the mutual coupling coefficients as a function of coupler rotations. We formulate a new problem to maximize the received signal-to-noise ratio (SNR) at the user by optimizing the 3D rotations of all couplers, subject to practical coupler rotation constraints. To tackle this nonconvex problem, we develop a spherical-cap conditional-gradient-based algorithm with cross-entropy-method initialization. Simulation results demonstrate that the proposed RCA system can significantly improve communication performance in comparison with benchmark schemes, while requiring substantially fewer active antennas and RF chains.
	\end{abstract}
	
	
\section{Introduction}
The evolution of wireless networks is entering a new era with the upcoming sixth-generation (6G) wireless systems \cite{10858129,10054381}. The 6G is expected to support various emerging applications, such as intelligent robotics, edge artificial intelligence (AI), and the Internet of Everything (IoE). These applications impose stringent and diverse requirements on data rate, latency, reliability, connectivity density, and energy efficiency. However, these requirements may not be fully met by simply following existing wireless technology trends, such as deploying more base stations (BSs) and access points (APs), equipping BSs with more antennas, and migrating to higher frequency bands such as millimeter-wave (mmWave) and terahertz (THz) bands \cite{9064545,9133130,exl,9724202, renwang1,10143420}. These approaches generally rely on a larger number of active antennas and radio-frequency (RF) chains to enhance network coverage, capacity, and beamforming gain. As a result, higher data rates are often achieved at the expense of larger installation space, higher hardware cost, higher power consumption, and more complicated signal processing. This issue becomes even more pronounced in high-frequency systems, where more RF chains are needed to compensate for severe propagation loss, and each RF chain is usually more expensive due to increased RF front-end loss and more stringent hardware requirements. Moreover, since the antennas in conventional multiple-input multiple-output (MIMO) systems are deployed at fixed positions and fixed orientations, the resulting wireless channels are mainly determined by the propagation environment and cannot be proactively reconfigured according to user distribution and environmental variations. Consequently, traditional techniques can only compensate for channel fading by using modulation, coding, diversity, power control, and beamforming, but they still have limited control over the largely random wireless channels \cite{9374451, 9903389,Larsson2014Massive}. In view of the above issues, it is desirable to develop new wireless system architectures that can achieve sustainable capacity growth with low cost, low complexity, and low power consumption.

To this end, flexible coupler antenna (FCA) has recently emerged as a promising new technique for cost-effective wireless channel reconfiguration \cite{shao2026coupler,FCATWC,FC1,FCjstsp}. Generally speaking, an FCA consists of one fixed-position active antenna connected to an RF chain and multiple low-cost passive couplers placed near the active antenna. The passive couplers can be independently adjusted in terms of their positions and/or rotations. The couplers are not connected to dedicated RF chains, but are excited through near-field electromagnetic (EM) mutual coupling with the active antenna and reradiate through the induced currents. Existing studies on FCA with flexible coupler positions have demonstrated the potential of FCA in different scenarios. In \cite{shao2026coupler}, coupler position optimization and centralized and distributed channel estimation were investigated for FCA-aided multiuser communication. In \cite{FCATWC}, an FCA-aided point-to-point communication system was studied, where passive coupler positions are optimized based on a block-coordinate conditional gradient method. In \cite{FCjstsp}, an FCA array was applied to low-altitude wireless communication, and an efficient parallel optimization method for coupler positioning and a coupler-on/off-based low-overhead channel estimation scheme were developed. In \cite{FC}, a dual-motion FCA is proposed, where passive couplers translate on the wavelength scale to enable mechanical beamforming, while the active antenna slides along a rail over larger distances to reconfigure the transceiver geometry. The work in \cite{FC1} provided an overview of FCA architectures, implementations, and applications in wireless networks. In addition, efficient iterative algorithms were developed in \cite{DuFCAload} for joint coupler-position and tunable-load optimization in FCA systems. Nevertheless, existing FCA systems still face limitations in terms of spatial flexibility and performance enhancement, since the existing works adjust only the positions of passive couplers while keeping their orientations fixed.

\begin{figure*}[t!]
	\centering
	\setlength{\abovecaptionskip}{0.cm}
	\includegraphics[width=5.9in]{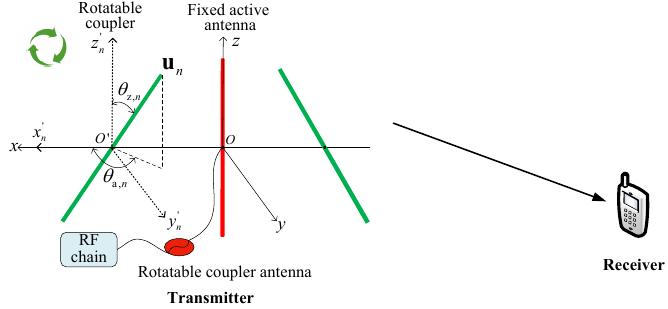}
	\caption{Proposed rotatable coupler antenna-aided point-to-point communication system.}
	\label{rca}
	\vspace{-0.39cm}
\end{figure*}

In this paper, we propose a new antenna structure, termed rotatable coupler antenna (RCA). As shown in Fig.~\ref{rca}, the proposed RCA consists of one fixed active antenna and multiple passive couplers with fixed centers. Each passive coupler can independently rotate to adjust its three-dimensional (3D) orientation (controlled by an attached compact rotary actuator), thereby allowing the passive couplers to collaboratively change the wireless propagation channel. Since the passive couplers are not connected to dedicated RF chains, they are excited by the fixed active antenna through near-field mutual coupling. Mutual coupling, i.e., the EM interaction between antennas, has traditionally been regarded as detrimental in antenna array design, since it may reduce antenna efficiency and complicate signal processing \cite{10500503,11176921}. To mitigate such coupling effects and reduce spatial correlation, typical antenna arrays maintain inter-element distances on the order of half a wavelength or larger, which makes their integration into compact wireless devices challenging. In contrast to this traditional view, the proposed RCA exploits mutual coupling as a useful design resource for improving system performance. Specifically, by controlling the rotations of passive couplers with directional element responses, the RCA can jointly reshape the effective wireless channel, the mutual impedance matrix, and the induced-current response. Accordingly, by properly adjusting the rotations of all passive couplers, the RCA can enhance the desired signal power at the intended receiver and alleviate destructive signal combining, thereby improving communication performance.

On the other hand, from the implementation perspective, the proposed RCA also possesses appealing advantages for practical deployment. Since the active antenna and its RF chain remain fixed, RCA only requires local rotation control of low-cost passive couplers around their fixed centers. This structure avoids the movement of RF-fed active antennas and does not require additional RF chains for the passive couplers, thereby reducing hardware cost, mechanical-control complexity, and power consumption. Moreover, the coupler rotation module can be implemented using compact rotary actuators, such as micro-electromechanical systems (MEMS)-based torsional actuators \cite{kurmendra2021review}. These features make RCA a promising solution for future wireless networks, particularly for compact and power-constrained devices with stringent size, weight, and power constraints, such as mobile terminals, unmanned aerial vehicles, and low-altitude communication platforms.

It is worth noting that the proposed RCA differs significantly from existing six-dimensional movable antenna (6DMA) systems \cite{shao20246d,6dma_dis,10945745,6DMA_JSTSP}. In 6DMA systems, active antennas are translated and/or rotated together with their RF chains to adapt to the spatial channel distribution \cite{li2025ai,liu2024uav,wang20256d, 10843383, 10891142,  wen, 10989638, near,10896748,shao2025tutorial, jiang2025statistical,6dmasensing,10906511,IPA}. In contrast, the proposed RCA keeps the active antenna and its RF chain fixed, and only rotates passive couplers around their fixed centers without adding dedicated RF chains. Therefore, RCA can cost-effectively reconfigure the wireless channel by changing the orientation-dependent directional responses of couplers. However, the modeling, performance analysis, and rotation optimization of RCA-enhanced wireless communication systems remain largely unexplored, which motivates this work. Here, we study an RCA-enhanced point-to-point communication system, where one RCA is deployed at the transmitter to serve a single user equipped with a fixed antenna. The main contributions of this work are summarized as follows:
\begin{itemize}
\item 
We propose a new RCA structure, where multiple passive couplers are placed near a fixed active antenna, while each coupler can independently rotate to adjust its 3D orientation. By exploiting orientation-dependent EM mutual coupling, the proposed RCA enhances wireless communication performance without requiring additional RF chains for the passive couplers. Based on the multi-port circuit theory framework, we establish the channel model and derive the mechanical beamforming vector of the passive couplers as explicit functions of the coupler rotations.
			
	\item  
Based on the established models, we formulate an optimization problem to maximize the received signal-to-noise ratio (SNR) at the user by designing the 3D rotations of the couplers, subject to practical rotation-range and physical non-intersection constraints. To solve this nonconvex problem, we develop a spherical-cap conditional-gradient-based algorithm, where a codebook-aided initialization first finds a high-quality feasible rotation axis matrix and the subsequent continuous refinement updates the coupler rotations while maintaining physical feasibility.
	
\item
Simulation results validate the effectiveness of the proposed RCA system and demonstrate that it can significantly improve communication performance compared with various benchmark schemes. The proposed RCA achieves a higher achievable rate than the fixed-geometry active antenna array while requiring only one transmit RF chain. Furthermore, the performance advantages of RCA become more pronounced with more couplers, a wider rotation range, and richer multipath propagation, which confirms the benefit of coupler-rotation-enabled wireless channel reconfigurability.
\end{itemize}

The remainder of this paper is organized as follows. Section~II describes the system model, including the channel model, coupler rotation constraints, and mechanical beamforming of the RCA. Section~III formulates the optimization problem for maximizing the received SNR through coupler rotation design. Section~IV presents the proposed optimization algorithm for solving this problem. Section~V provides numerical results and corresponding discussions. Finally, Section~VI concludes this study.

\emph{Notations.}
Boldface uppercase and lowercase letters denote matrices and vectors, respectively. Symbols
$(\cdot)^{\mathrm H}$ and $(\cdot)^{\mathrm T}$ denote conjugate transpose and transpose, respectively.
For scalar $a$, $|a|$ denotes its magnitude.
For vector $\mathbf{a}$, $\|\mathbf{a}\|_2$ denotes the $\ell_{2}$ norm. Symbol
$\mathrm{diag}(\mathbf{x})$ denotes a diagonal matrix whose diagonal is $\mathbf{x}$,
$[\mathbf{a}]_{j}$ denotes the $j$th entry of a vector $\mathbf{a}$, $[\mathbf{A}]_{i,j}$ denotes the $(i,j)$ entry of a matrix $\mathbf{A}$,
$\mathbb{R}$ and $\mathbb{C}$ denote the real and complex fields, respectively,
$\mathcal{CN}(\mu,\sigma^{2})$ denotes a complex Gaussian distribution with mean $\mu$ and variance $\sigma^{2}$,
$\Re\{\cdot\}$ and $\Im\{\cdot\}$ denote the real and imaginary parts of a complex quantity, respectively, $\mathcal{O}(\cdot)$ denotes the big O notation, $\odot$ denotes the Hadamard product, $\mathbb{E}[\cdot]$ denotes statistical expectation, and $\mathbf I_M$ denotes the $M\times M$ identity matrix.

\section{System Model}
As shown in Fig.~\ref{rca}, we consider a point-to-point communication system, where a transmitter is equipped with an RCA and a receiver has a fixed-position/orientation isotropic antenna. The RCA consists of one fixed active antenna and $N$ rotatable couplers. Each coupler's rotation can be independently adjusted in 3D space by mechanical means. The active antenna is connected to a single RF chain, whereas the passive couplers are not connected to dedicated RF chains. Instead, the passive couplers are excited through near-field EM mutual coupling with the active antenna and reradiate through the induced currents. For geometric and mutual-impedance modeling, the couplers and the active antenna are modeled as straight thin-wire dipoles with identical length $D$ and radius $a$. 

\begin{figure}[t!]
	\centering
	\setlength{\abovecaptionskip}{0.cm}
	\includegraphics[width=3.5in]{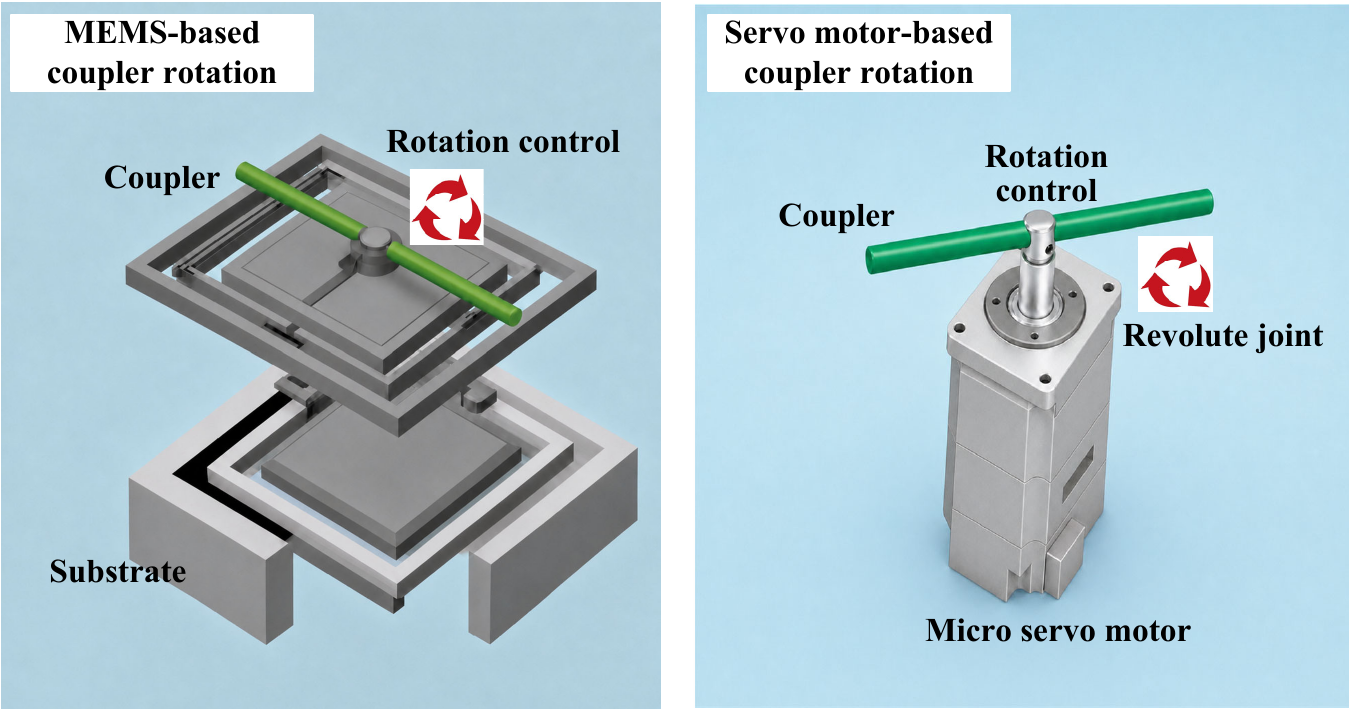}
	\caption{Hardware implementation of the proposed RCA.}
	\label{motor}
\end{figure}

To further demonstrate the practical feasibility of the proposed RCA, Fig.~\ref{motor} presents two possible hardware implementations for realizing the rotatable coupler in Fig.~\ref{rca}. Specifically, the proposed RCA can be realized by integrating a coupler rotation module with the conventional communication module. In the coupler rotation module, a central processing unit (CPU) is deployed at the RCA-equipped transmitter for local control. Based on the channel information, the CPU computes the desired coupler rotations and sends control commands to compact rotary actuators, such as MEMS-based torsional actuators or miniature servo motors, mounted at the fixed centers of the couplers, as shown in Fig.~\ref{motor}. Thus, each coupler changes only its rotation while its center position remains fixed. Note that MEMS-based actuators generally operate with milliwatt-level power consumption and millisecond-scale response times~\cite{yang2025low}.

\subsection{Channel Model}
As shown in Fig.~\ref{rca}, we establish a global Cartesian coordinate system (CCS) $o$-$xyz$, whose origin $o$ is located at the center-fed port of the fixed active antenna. Unless otherwise specified, all position vectors and direction vectors are represented in this global CCS. Thus, the center position of the active antenna is given by
\begin{align}
	\mathbf p_0=[0,0,0]^{\mathrm T}.
	\label{eq:p0}
\end{align}
The fixed active antenna is aligned with the positive $z$-axis. Its axis vector, which specifies the antenna orientation in the global CCS, is given by
\begin{align}
	\mathbf u_0=[0,0,1]^{\mathrm T}.
	\label{eq:u0}
\end{align}
The centers of all couplers are fixed on the $x$-axis. For the $n$-th passive coupler, its center position is denoted by
\begin{align}
	\mathbf p_n=[x_n,0,0]^{\mathrm T},  n\in\mathcal N,
	\label{eq:pn}
\end{align}
where $\mathcal N\triangleq\{1,2,\ldots,N\}$ and $x_n$ is predetermined. Hence, the centers of the active antenna and all passive couplers are fixed on the $x$-axis.
We denote the index set of all transmitter antenna elements as $\mathcal I\triangleq\{0\}\cup\mathcal N$, where index $0$ corresponds to the active antenna and the indices in $\mathcal N$ correspond to the passive couplers.

With these fixed centers, the proposed RCA model allows each coupler to adjust its axis orientation in 3D space. To describe such rotations, a reference local CCS $o_n'$-$x_n'y_n'z_n'$ is attached to the center of the $n$-th coupler, where local origin $o_n'$ is located at $\mathbf p_n$ and the local axes are parallel to the corresponding axes of the global CCS (see Fig.~\ref{rca}). The local CCS is used only to define the rotation angles at the coupler center and remains fixed during coupler rotation. Since each coupler is modeled as a straight thin-wire dipole, its actual orientation is fully specified by the rotation axis vector expressed in the global CCS. 

Let $\mathbf u_n\in\mathbb R^{3\times1}$ denote the rotation axis vector of the $n$-th coupler. As illustrated in Fig.~\ref{rca}, $\mathbf u_n$ is parameterized by the zenith angle $\theta_{{\mathrm z},n}$ and the azimuth angle $\theta_{{\mathrm a},n}$. Specifically, $\theta_{{\mathrm z},n}$ denotes the angle between $\mathbf u_n$ and the local $z_n'$-axis, while $\theta_{{\mathrm a},n}$ denotes the angle from the local $x_n'$-axis to the projection of $\mathbf u_n$ onto the local $x_n'$-$y_n'$ plane. Therefore, the rotation axis vector of the $n$-th coupler can be expressed as
\begin{align}
	\mathbf u_n(\theta_{{\mathrm{z}},n},\theta_{{\mathrm{a}},n})
	=
	\begin{bmatrix}
		\sin\theta_{{\mathrm{z}},n}\cos\theta_{{\mathrm{a}},n}\\
		\sin\theta_{{\mathrm{z}},n}\sin\theta_{{\mathrm{a}},n}\\
		\cos\theta_{{\mathrm{z}},n}
	\end{bmatrix},
	n\in\mathcal N.
	\label{eq:un_3d}
\end{align}
By construction, we have $\|\mathbf u_n\|_2=1$. Accordingly, the rotation axis matrix of all couplers is denoted as
\begin{align}
	\mathbf U
	=
	[\mathbf u_1,\mathbf u_2,\ldots,\mathbf u_N]
	\in\mathbb R^{3\times N}.
	\label{eq:U_def}
\end{align}

For the baseband channel model, the active antenna is treated as an omnidirectional element, whereas the couplers are modeled with directional element responses. We consider a narrowband far-field geometric multipath channel with $L$ propagation paths. Let $\mathcal L\triangleq\{1,2,\ldots,L\}$ denote the set of channel paths. For path $\ell\in\mathcal L$, its departure direction from the transmitter is represented by unit vector
\begin{align}
	\mathbf f_\ell
	=
	[\sin\psi_\ell\cos\phi_\ell,\;
	\sin\psi_\ell\sin\phi_\ell,\;
	\cos\psi_\ell]^{\mathrm T},
	\label{eq:fl}
\end{align}
where $\psi_\ell\in[0,\pi]$ and $\phi_\ell\in[-\pi,\pi]$ denote the zenith and azimuth angles of path $\ell$, respectively.

Accordingly, the steering vector associated with the fixed antenna and coupler centers is given by
\begin{align}
	\mathbf a_\ell
	=
	\left[
	1,\;
	e^{\mathrm j k\mathbf f_\ell^{\mathrm T}\mathbf p_1},
	\ldots,
	e^{\mathrm j k\mathbf f_\ell^{\mathrm T}\mathbf p_N}
	\right]^{\mathrm T}
	\in\mathbb C^{(N+1)\times1},
	\label{eq:al}
\end{align}
where $\lambda$ denotes the wavelength, and $k=2\pi/\lambda$ denotes the wavenumber.
Since all coupler and antenna centers are fixed, steering vector $\mathbf a_\ell$ is independent of rotation axis matrix $\mathbf U$.

For the $n$-th coupler and path $\ell$, the direction cosine between the rotation axis vector and the path direction is denoted as
\begin{align}
	\xi_{n,\ell}(\mathbf u_n)
	\triangleq
	\mathbf u_n^{\mathrm T}\mathbf f_\ell,
	~~
	n\in\mathcal N,\quad \ell\in\mathcal L.
	\label{eq:xi_nl}
\end{align}

Then, the unnormalized directional field response of the $n$-th thin-wire-shaped coupler along path $\ell$ is given by \cite{balanis2016antenna,10829613}
\begin{align}
	\widetilde e_{n,\ell}(\mathbf u_n)
	=
	\frac{
		\cos\!\left(\frac{kD}{2}\xi_{n,\ell}(\mathbf u_n)\right)
		-
		\cos\!\left(\frac{kD}{2}\right)
	}{
		\sqrt{1-\xi_{n,\ell}^{2}(\mathbf u_n)}
	},
	\label{eq:edip_unnormalized}
\end{align}
for $n\in\mathcal N$ and $\ell\in\mathcal L$. When $\xi_{n,\ell}(\mathbf u_n)=\pm1$, continuous extension $\widetilde e_{n,\ell}(\mathbf u_n)=0$ is used, which is obtained from the limit of \eqref{eq:edip_unnormalized} as the observation direction approaches the antenna axis.

To avoid artificial power variations caused only by pattern normalization, the thin-wire antenna pattern is normalized over the full 3D angular domain. The normalization coefficient is given by
\begin{align}
	c_{\mathrm{dip}}
	&=
	\Bigg(
	\frac{1}{4\pi}
	\int_{0}^{\pi}
	\int_{-\pi}^{\pi}
	\left|
	\frac{
		\cos\!\left(
		\frac{kD}{2}
		\mathbf u_0^{\mathrm T}\mathbf f(\zeta,\omega)
		\right)
		-
		\cos\!\left(\frac{kD}{2}\right)
	}{
		\sqrt{
			1-
			\left(
			\mathbf u_0^{\mathrm T}\mathbf f(\zeta,\omega)
			\right)^2
		}
	}
	\right|^2
	\notag\\
	&\qquad
	\times
	\sin\zeta\, d\omega\, d\zeta
	\Bigg)^{-1/2},
	\label{eq:cdip}
\end{align}
where $\mathbf f(\zeta,\omega)
=
[\sin\zeta\cos\omega,\;
\sin\zeta\sin\omega,\;
\cos\zeta]^{\mathrm T}$ denotes a generic unit observation direction, with $\zeta\in[0,\pi]$ and $\omega\in[-\pi,\pi]$ denoting the zenith and azimuth angles, respectively. Since the integration is performed over the entire 3D angular domain, the normalization coefficient $c_{\mathrm{dip}}$ is independent of the coupler rotation.

Then, the normalized element radiation pattern of the $n$-th coupler along path $\ell$ is obtained as
\begin{align}
	e_{n,\ell}(\mathbf u_n)
	=
	c_{\mathrm{dip}}
	\widetilde e_{n,\ell}(\mathbf u_n),
	~~
	n\in\mathcal N,\quad \ell\in\mathcal L.
	\label{eq:enl_dip}
\end{align}

Moreover, since the active antenna is modeled as an omnidirectional element, its element response is given by
\begin{align}
	e_{0,\ell}=1,
	\qquad
	\ell\in\mathcal L.
	\label{eq:e0l_omni}
\end{align}

Accordingly, the overall radiation pattern vector of path $\ell$ is given by
\begin{align}
	\mathbf e_\ell(\mathbf U)
	=
	\sqrt{\frac{\eta}{\pi}}
	[1,
	e_{1,\ell}(\mathbf u_1),
	\ldots,
	e_{N,\ell}(\mathbf u_N)]^{\mathrm T}
	\in\mathbb C^{(N+1)\times1},
	\label{eq:el}
\end{align}
where $\eta$ denotes the wave impedance of the propagation medium. Combining the steering vector in \eqref{eq:al} and the radiation pattern vector in \eqref{eq:el}, the effective channel vector of path $\ell$ is expressed as
\begin{align}
	\mathbf g_\ell(\mathbf U)
	=
	\mathbf a_\ell\odot \mathbf e_\ell(\mathbf U)
	\in\mathbb C^{(N+1)\times1}.
	\label{eq:gl}
\end{align}

Therefore, the overall multipath channel vector from the RCA transmitter to the receiver, $\mathbf h(\mathbf U)\in\mathbb C^{(N+1)\times1}$, is given by
\begin{align}
	\mathbf h(\mathbf U)
	&=
	\sum_{\ell=1}^{L}\gamma_\ell\mathbf g_\ell(\mathbf U)\nonumber\\
	&=
	\mathbf G(\mathbf U)\boldsymbol\gamma,
	\label{eq:hU}
\end{align}
where $\gamma_\ell$ denotes the complex gain of path $\ell$,
$\mathbf G(\mathbf U)
=
[\mathbf g_1(\mathbf U),\mathbf g_2(\mathbf U),\ldots,\mathbf g_L(\mathbf U)]
\in\mathbb C^{(N+1)\times L}$, and
$\boldsymbol\gamma
=
[\gamma_1,\gamma_2,\ldots,\gamma_L]^{\mathrm T}
\in\mathbb C^{L\times1}$.

\subsection{Coupler Rotation Constraints}
Next, we introduce two practical constraints for the 3D rotations of the couplers.
\subsubsection{Rotation Range Constraints}
We limit the feasible 3D rotation range of each coupler by imposing a unit-norm constraint and a maximum zenith-rotation constraint.

First, since $\mathbf u_n$ represents only the orientation of the $n$-th coupler and is independent of the physical antenna length, it is defined as a unit vector, i.e.,
\begin{align}
	\|\mathbf u_n\|_2=1,~ n\in\mathcal N.
	\label{eq:unit_un}
\end{align} 

Second, we limit the zenith and azimuth rotation angles as
\begin{align}
	0\le \theta_{{\mathrm{z}},n}\le \theta_{\max},
	~
	-\pi\le \theta_{{\mathrm{a}},n}<\pi,
	~ n\in\mathcal N,
	\label{eq:angle_constraints}
\end{align}
where $\theta_{\max}$ denotes the maximum allowable rotation angle from reference axis $\mathbf u_0$. Constraint~\eqref{eq:angle_constraints} limits the zenith rotation of each passive coupler within $\theta_{\max}$, while allowing full azimuth rotation. Under the angular parameterization in \eqref{eq:un_3d}, the maximum zenith-rotation constraint can be equivalently written in vector form as
\begin{align}
	\mathbf u_0^{\mathrm T}\mathbf u_n
	\ge
	\cos\theta_{\max},
	~ n\in\mathcal N.
	\label{eq:cone_un}
\end{align}
Since both $\mathbf u_0$ and $\mathbf u_n$ are unit vectors, $\mathbf u_0^{\mathrm T}\mathbf u_n$ equals the cosine of the angle between reference axis $\mathbf u_0$ and rotation axis vector $\mathbf u_n$. Thus, constraint~\eqref{eq:cone_un} restricts $\mathbf u_n$ to lie inside a 3D cone with axis $\mathbf u_0$ and half-angle $\theta_{\max}$.

\subsubsection{Non-Intersection Constraints}
\begin{figure}[t!]
	\centering
	\setlength{\abovecaptionskip}{0.cm}
	\includegraphics[width=3.5in]{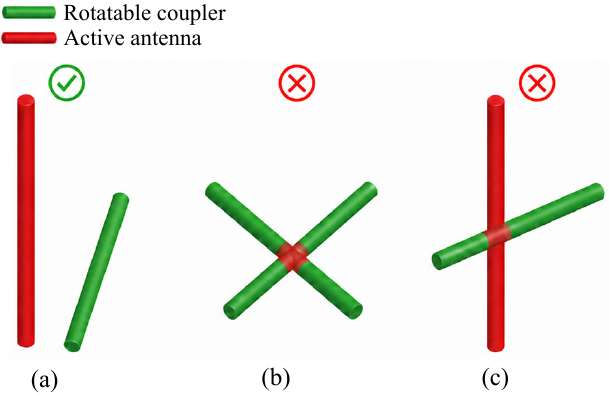}
	\caption{Illustration of the physical non-intersection constraint for the proposed RCA. (a) Feasible configuration. (b) Infeasible coupler-coupler overlap. (c) Infeasible active-coupler overlap.}
	\label{constraints}
\end{figure}
Since all couplers and active antenna elements are modeled as cylindrical thin-wire dipoles with length $D$ and radius $a$, their physical overlap must be avoided during 3D rotation (see Fig.~\ref{constraints}). To this end, we explicitly impose a non-intersection constraint based on the minimum distance between the axis segments of any two antenna elements.

Using the global CCS and the rotation axis vectors defined above, each thin-wire antenna element can be described by a scalar local axial parameter and its actual axis direction. Specifically, for antenna element $i\in\mathcal I$, let $s\in[-D/2,D/2]$ denote the signed distance from the element center along its own axis. Thus, $s=0$ corresponds to the element center, while $s=D/2$ and $s=-D/2$ correspond to the two endpoints of the antenna element along the directions $\mathbf u_i$ and $-\mathbf u_i$, respectively. Then, the point on antenna element $i$ with axial parameter $s$ is given by
\begin{align}
	\mathbf r_i(s;\mathbf u_i)
	=
	\mathbf p_i+s\mathbf u_i,
	~ i\in\mathcal I,
	\label{eq:ri_s}
\end{align}
where $\mathbf p_i$ is the center position of antenna element $i$, and $\mathbf u_i$ denotes its rotation axis vector in the global CCS. Therefore, for any two distinct antenna elements $i,j\in\mathcal I$ with $i<j$, the minimum distance between their antenna-axis segments is obtained as
\begin{align}
	d_{i,j}(\mathbf U)
	\!	=\!
	\min_{s_i,s_j\in[-D/2,D/2]}
	\left\|
	\mathbf r_i(s_i;\mathbf u_i)
	-
	\mathbf r_j(s_j;\mathbf u_j)
	\right\|_2 .
	\label{eq:dij}
\end{align}
To prevent physical overlap between any two cylindrical antenna elements, their minimum axis-segment distance must be no smaller than the antenna diameter, i.e., $2a$. Therefore, the non-intersection constraints are given by
\begin{align}
	d_{i,j}(\mathbf U)\ge 2a,
	~ \forall i,j\in\mathcal I,\ i<j.
	\label{eq:non_intersection}
\end{align}

\subsection{Mechanical Beamforming Vector}
The EM mutual coupling among the active antenna and all passive couplers is characterized by the transmit impedance matrix. With port ordering $\mathcal I=\{0\}\cup\mathcal N$, this matrix is given by
\begin{align}
	\mathbf Z_{\mathrm{TX}}(\mathbf U)
	=
	\begin{bmatrix}
		z_{\mathrm{s}} & \bar{\mathbf z}^{\mathrm T}(\mathbf U)\\
		\bar{\mathbf z}(\mathbf U) & \mathbf Z_{\mathrm E}(\mathbf U)
	\end{bmatrix}
	\in\mathbb C^{(N+1)\times(N+1)},
	\label{eq:Ztx_block}
\end{align}
where $z_{\mathrm{s}}$ is the common self-impedance of each thin-wire antenna, $\bar{\mathbf z}(\mathbf U)\in\mathbb C^{N\times1}$ collects the mutual impedances between the active antenna and the couplers, and $\mathbf Z_{\mathrm E}(\mathbf U)\in\mathbb C^{N\times N}$ is the impedance submatrix among all couplers with diagonal entries $z_{\mathrm{s}}$. The detailed computation of $\mathbf Z_{\mathrm{TX}}(\mathbf U)$ is provided in the Appendix.

Let $i_0$ denote the current coefficient at the active antenna. For coupler $j\in\mathcal N$, let $i_j$ denote its current coefficient. Let $v_0$ and $v_j$ denote the corresponding port voltages at the active antenna and coupler $j$, respectively. The voltage vector and current-coefficient vector of all couplers are denoted by
\begin{align}
	\mathbf v_{\mathrm E}
	=
	[v_1,\ldots,v_N]^{\mathrm T}
	\in\mathbb C^{N\times1},
	\label{eq:vE}
\end{align}
and
\begin{align}
	\mathbf i_{\mathrm E}
	=
	[i_1,\ldots,i_N]^{\mathrm T}
	\in\mathbb C^{N\times1}.
	\label{eq:iE}
\end{align}
Accordingly, the stacked voltage vector and current-coefficient vector of all transmitter ports are given by
\begin{align}
	\mathbf v_{\mathrm{TX}}
	=
	[v_0,\mathbf v_{\mathrm E}^{\mathrm T}]^{\mathrm T}
	\in\mathbb C^{(N+1)\times1},
	\label{eq:vTX}
\end{align}
and
\begin{align}
	\mathbf i_{\mathrm{TX}}
	=
	[i_0,\mathbf i_{\mathrm E}^{\mathrm T}]^{\mathrm T}
	\in\mathbb C^{(N+1)\times1}.
	\label{eq:iTX}
\end{align}

Let the coupler load impedance matrix be 
\begin{align}
	\mathbf X
	=
	\mathrm{diag}\{X_{\mathrm C,1},X_{\mathrm C,2},\ldots,X_{\mathrm C,N}\}
	\in\mathbb C^{N\times N},
	\label{eq:Xload}
\end{align}
where $X_{\mathrm C,n}$ denotes the load impedance connected to the $n$-th coupler. Since the couplers are terminated by load impedances and are not connected to dedicated RF chains, their port voltages are constrained by the induced currents and the load impedances. Under the standard port convention, the coupler port voltages satisfy
\begin{align}
	\mathbf v_{\mathrm E}
	=
	-\mathbf X\mathbf i_{\mathrm E}\chi,
	\label{eq:vE_load}
\end{align}
where $\chi\in\mathbb C$ denotes the transmit information symbol and the negative sign indicates that the load voltage has the opposite reference direction to the antenna-port voltage.

Using the unilateral approximation, the transmit-port voltage vector is obtained as
\begin{align}
	\mathbf v_{\mathrm{TX}}
	\approx
	\mathbf Z_{\mathrm{TX}}(\mathbf U)\mathbf i_{\mathrm{TX}}\chi.
	\label{eq:vTX_ZTX}
\end{align}
This approximation neglects the feedback from the far-field receiver to the transmitter. Thus, the transmitter-port voltages are determined by the transmit impedance matrix and the transmitter-port current coefficients multiplied by $\chi$. Extracting the coupler rows of \eqref{eq:vTX_ZTX} gives
\begin{align}
	\mathbf v_{\mathrm E}
	=
	\bar{\mathbf z}(\mathbf U)i_0\chi
	+
	\mathbf Z_{\mathrm E}(\mathbf U)\mathbf i_{\mathrm E}\chi,
	\label{eq:vE_extract}
\end{align}
where the first term is induced by the active-antenna current through the active-to-coupler mutual impedances, while the second term is caused by the self- and mutual-impedance interactions among the couplers.

Combining \eqref{eq:vE_load} and \eqref{eq:vE_extract}, and canceling the common transmit symbol $\chi$, we obtain
\begin{align}
	\mathbf i_{\mathrm E}
	=
	-
	\big(\mathbf Z_{\mathrm E}(\mathbf U)+\mathbf X\big)^{-1}
	\bar{\mathbf z}(\mathbf U)i_0.
	\label{eq:iE_solution}
\end{align}
Equation~\eqref{eq:iE_solution} shows that the coupler currents are not independently generated by RF chains. Instead, they are induced by the active-antenna current through near-field EM mutual coupling and depend on the coupler rotations through $\bar{\mathbf z}(\mathbf U)$ and $\mathbf Z_{\mathrm E}(\mathbf U)$.

For compactness, the mechanical beamforming vector induced by the couplers is denoted by
\begin{align}
	\mathbf w(\mathbf U)
	\triangleq
	\big(\mathbf Z_{\mathrm E}(\mathbf U)+\mathbf X\big)^{-1}
	\bar{\mathbf z}(\mathbf U)
	\in\mathbb C^{N\times1}.
	\label{eq:w_def}
\end{align}
Then, the coupler current-coefficient vector can be written as
\begin{align}
	\mathbf i_{\mathrm E}
	=
	-\mathbf w(\mathbf U)i_0.
	\label{eq:iE_w}
\end{align}
Thus, the coupler rotations modify $\mathbf w(\mathbf U)$ through the rotation-dependent impedance matrix, thereby shaping the induced currents without adding RF chains.

Finally, the overall transmit current-coefficient vector can be expressed as
\begin{align}
	\mathbf i_{\mathrm{TX}}
	=
	\mathbf w_{\mathrm e}(\mathbf U)i_0,
	\label{eq:iTX_we}
\end{align}
where
\begin{align}
	\mathbf w_{\mathrm e}(\mathbf U)
	=
	[1,-\mathbf w^{\mathrm T}(\mathbf U)]^{\mathrm T}
	\in\mathbb C^{(N+1)\times1},
	\label{eq:we_def}
\end{align}
denotes the overall mechanical beamforming vector of the RCA.

\section{Problem Formulation}

Based on the channel vector and the mechanical beamforming vector derived in the previous section, the received baseband signal is expressed as
\begin{align}
	y
	=
	\mathbf h^{\mathrm T}(\mathbf U)\mathbf i_{\mathrm{TX}}\chi+\varpi 
	=
	\mathbf h^{\mathrm T}(\mathbf U)\mathbf w_{\mathrm e}(\mathbf U)i_0\chi+\varpi ,
	\label{eq:y}
\end{align}
where $\chi$ is the transmit information symbol satisfying $\mathbb E[|\chi|^2]=1$, and $\varpi \sim\mathcal{CN}(0,\sigma^2)$ denotes the complex Gaussian noise with zero mean and variance $\sigma^2$, due to the additive white Gaussian noise from the channel. Accordingly, the received SNR is calculated as
\begin{align}
	r(\mathbf U)
	=
	\frac{
		|i_0|^2
		\left|
		\mathbf h^{\mathrm T}(\mathbf U)\mathbf w_{\mathrm e}(\mathbf U)
		\right|^2
	}{
		\sigma^2
	}.
	\label{eq:snr_raw}
\end{align}

The total transmit power radiated by the RCA should account for the EM mutual coupling among all transmitter ports. Based on the multi-port circuit model, the radiated transmit power is given by
\begin{subequations}
	\begin{align}
		P_{\mathrm{TX}}(\mathbf U)
		&=
		\mathbb E
		\left[
		\Re\left\{
		(\mathbf i_{\mathrm{TX}}\chi)^{\mathrm H}\mathbf v_{\mathrm{TX}}
		\right\}
		\right]
		\label{eq:ptx_1}\\
		&=
		\mathbb E
		\left[
		\Re\left\{
		(\mathbf i_{\mathrm{TX}}\chi)^{\mathrm H}
		\mathbf Z_{\mathrm{TX}}(\mathbf U)\mathbf i_{\mathrm{TX}}\chi
		\right\}
		\right]
		\label{eq:ptx_2}\\
		&=
		\mathbf i_{\mathrm{TX}}^{\mathrm H}
		\Re\{\mathbf Z_{\mathrm{TX}}(\mathbf U)\}
		\mathbf i_{\mathrm{TX}}
		\label{eq:ptx_3}\\
		&=
		|i_0|^2
		\mathbf w_{\mathrm e}^{\mathrm H}(\mathbf U)
		\Re\{\mathbf Z_{\mathrm{TX}}(\mathbf U)\}
		\mathbf w_{\mathrm e}(\mathbf U),
		\label{eq:ptx_4}
	\end{align}
\end{subequations}
where \eqref{eq:ptx_2} follows from the unilateral approximation in \eqref{eq:vTX_ZTX}, \eqref{eq:ptx_3} follows from $\mathbb E[|\chi|^2]=1$, and \eqref{eq:ptx_4} follows from $\mathbf i_{\mathrm{TX}}=\mathbf w_{\mathrm e}(\mathbf U)i_0$.

Given transmit power budget $P$, we choose the active-antenna current amplitude such that the radiated transmit power satisfies $P_{\mathrm{TX}}(\mathbf U)=P$. From \eqref{eq:ptx_4}, this yields
\begin{align}
	|i_0|^2
	=
	\frac{
		P
	}{
		\mathbf w_{\mathrm e}^{\mathrm H}(\mathbf U)
		\Re\{\mathbf Z_{\mathrm{TX}}(\mathbf U)\}
		\mathbf w_{\mathrm e}(\mathbf U)
	}.
	\label{eq:i0_power}
\end{align}
Substituting \eqref{eq:i0_power} into \eqref{eq:snr_raw}, the received SNR can be rewritten as
\begin{align}
	r(\mathbf U)
	=
	\frac{
		P
		\left|
		\mathbf h^{\mathrm T}(\mathbf U)\mathbf w_{\mathrm e}(\mathbf U)
		\right|^2
	}{
		\sigma^2
		\mathbf w_{\mathrm e}^{\mathrm H}(\mathbf U)
		\Re\{\mathbf Z_{\mathrm{TX}}(\mathbf U)\}
		\mathbf w_{\mathrm e}(\mathbf U)
	}.
	\label{eq:snr_final}
\end{align}

We aim to maximize the received SNR by optimizing the 3D rotations of all couplers. Since $P/\sigma^2$ is independent of $\mathbf U$, maximizing \eqref{eq:snr_final} is equivalent to maximizing the normalized SNR gain. Therefore, the 3D coupler rotation optimization problem is formulated as
\begin{subequations}
	\begin{align}
		(\mathrm{P1}):\quad
		\max_{\mathbf U}\quad
		&
		\frac{
			\left|
			\mathbf h^{\mathrm T}(\mathbf U)\mathbf w_{\mathrm e}(\mathbf U)
			\right|^2
		}{
			\mathbf w_{\mathrm e}^{\mathrm H}(\mathbf U)
			\Re\{\mathbf Z_{\mathrm{TX}}(\mathbf U)\}
			\mathbf w_{\mathrm e}(\mathbf U)
		}
		\label{eq:P1_obj}\\
		\mathrm{s.t.}\quad
		&
		\|\mathbf u_n\|_2=1,
		~ n\in\mathcal N,
		\label{eq:P1_unit}\\
		&
		\mathbf u_0^{\mathrm T}\mathbf u_n\ge \cos\theta_{\max},
		~ n\in\mathcal N,
		\label{eq:P1_cone}\\
		&
		d_{i,j}(\mathbf U)\ge 2a,
		~
		\forall\, i,j\in\mathcal I,\ i<j,
		\label{eq:P1_nonintersection}
	\end{align}
\end{subequations}
where constraint~\eqref{eq:P1_unit} ensures that each $\mathbf u_n$ is a unit rotation axis vector. Constraint~\eqref{eq:P1_cone} limits the maximum zenith rotation of each coupler with respect to reference axis $\mathbf u_0$. Constraint~\eqref{eq:P1_nonintersection} prevents physical overlap between any two antenna elements during coupler rotation.

Problem~(P1) is highly non-convex due to the coupled dependence of the objective function on the coupler rotations. First, effective channel vector $\mathbf h(\mathbf U)$ depends nonlinearly on $\mathbf U$ through the rotation-dependent  antenna responses of the couplers. Second, the transmit impedance matrix $\mathbf Z_{\mathrm{TX}}(\mathbf U)$ depends nonlinearly on $\mathbf U$ through the non-coplanar mutual impedances computed in \eqref{eq:z_non_coplanar}. Third, overall mechanical beamforming vector $\mathbf w_{\mathrm e}(\mathbf U)$ is an implicit nonlinear function of $\mathbf U$. These coupled nonlinearities make it difficult to obtain the globally optimal coupler rotations in a closed form, which motivates the algorithm design in the next section.

\section{Coupler Rotation Optimization Algorithm}
In this section, we develop an efficient algorithm for solving problem~(P1). The main difficulty lies in the fact that the design variables are not unconstrained Euclidean variables. Each coupler is represented by a unit axis vector confined to a spherical cap, and all axis vectors must jointly satisfy the physical non-intersection constraints. To solve this constrained non-convex problem, we first derive a continuous spherical-cap conditional-gradient update with a closed-form linear oracle. Then, we use a codebook-aided cross-entropy method (CEM) initialization to provide a feasible starting point for the continuous refinement.

For notational convenience, we denote the objective function of problem~(P1) by
\begin{align}
	\Omega(\mathbf U)
	\triangleq
	\frac{
		\left|
		\mathbf h^{\mathrm T}(\mathbf U)
		\mathbf w_{\mathrm e}(\mathbf U)
		\right|^2
	}{
		\mathbf w_{\mathrm e}^{\mathrm H}(\mathbf U)
		\Re\{\mathbf Z_{\mathrm{TX}}(\mathbf U)\}
		\mathbf w_{\mathrm e}(\mathbf U)
	}.
	\label{eq:Omega_def}
\end{align}
Since $\ln(\cdot)$ is monotonically increasing, maximizing $\Omega(\mathbf U)$ is equivalent to maximizing
\begin{align}
	\Phi(\mathbf U)
	\triangleq
	\ln \Omega(\mathbf U).
	\label{eq:Phi_def}
\end{align}
The logarithmic objective in \eqref{eq:Phi_def} preserves the optimizer of problem~(P1) and improves numerical stability when the normalized SNR gain varies over a large dynamic range.

Following the standard definition of a spherical cap on the unit sphere~\cite{li2024spherical}, the individual feasible rotation set of each coupler is
\begin{align}
	\mathcal C_{\mathrm{cap}}
	\triangleq
	\left\{
	\mathbf x\in\mathbb R^{3}
	\,\middle|\,
	\|\mathbf x\|_2=1,\ 
	\mathbf u_0^{\mathrm T}\mathbf x\ge\cos\theta_{\max}
	\right\}.
	\label{eq:Ccap_RCA}
\end{align}
The unit-norm condition states that $\mathbf x$ is only an axis direction, whereas the inner-product condition limits the maximum zenith rotation from reference axis $\mathbf u_0$. By adding the pairwise non-intersection constraints, the feasible set of the rotation axis matrix becomes
\begin{align}
	\mathcal U
	\triangleq
	\left\{
	\mathbf U
	\,\middle|\,
	\begin{array}{l}
		\mathbf u_n\in\mathcal C_{\mathrm{cap}},\quad \forall n\in\mathcal N,\\
		d_{i,j}(\mathbf U)\ge 2a,\quad \forall i,j\in\mathcal I,\ i<j
	\end{array}
	\right\},
	\label{eq:URCA_feasible}
\end{align}
where the $n$-th column of $\mathbf U$ is $\mathbf u_n$.

Thus, (P1) is equivalently written as
\begin{align}
	\max_{\mathbf U\in\mathcal U}\quad
	\Phi(\mathbf U).
	\label{eq:P2}
\end{align}

\subsection{Spherical-Cap Conditional-Gradient Update}
Starting from a feasible matrix, $\mathbf U^{(0)}\in\mathcal U$, the proposed continuous refinement updates all coupler rotations iteratively. At iteration $t$, let
\begin{align}
	\mathbf U^{(t)}
	=
	[\mathbf u_1^{(t)},\mathbf u_2^{(t)},\ldots,\mathbf u_N^{(t)}]
	\in\mathcal U.
	\label{eq:Ut_def}
\end{align}
Since the mutual impedance matrix, the induced currents on the couplers, and the effective channel all depend nonlinearly on $\mathbf U$, deriving a closed-form expression for the gradient of $\Phi(\mathbf U)$ is analytically cumbersome. We therefore estimate a local first-order ascent model in the tangent space of the unit sphere. The spherical-cap geometry is then explicitly exploited by using a cap retraction to keep trial rotations within $\mathcal C_{\mathrm{cap}}$ and a cap-constrained linear oracle to select the best feasible search point under the local first-order model.

Specifically, the constants used in the spherical-cap geometry are 
\begin{align}
	c_{\theta}
	=
	\cos\theta_{\max},
	\qquad
	s_{\theta}
	=
	\sin\theta_{\max}.
	\label{eq:ctheta_stheta}
\end{align}
Inspired by projection-based retractions for constrained manifold optimization~\cite{absil2012projection}, we introduce a cap retraction mapping, $\mathcal R_{\mathrm{cap}}(\cdot)$, to ensure that trial rotations remain in $\mathcal C_{\mathrm{cap}}$. Specifically, for any $\mathbf y\in\mathbb R^3$, let $\bar{\mathbf y}=\mathbf y/\|\mathbf y\|_2$ when $\mathbf y\neq \mathbf 0$, and let
\begin{align}
	\mathbf y_{\perp}
	\triangleq
	(\mathbf I_3-\mathbf u_0\mathbf u_0^{\mathrm T})\mathbf y .
	\label{eq:y_perp_def}
\end{align}
The cap retraction is defined as
\begin{align}
	\mathcal R_{\mathrm{cap}}(\mathbf y)
	=
	\begin{cases}
		\mathbf u_0,
		&
		\mathbf y=\mathbf 0,
		\\[1mm]
		\bar{\mathbf y},
		&
		\mathbf y\ne\mathbf 0,\ 
		\bar{\mathbf y}^{\mathrm T}\mathbf u_0\ge c_{\theta},
		\\[1mm]
		c_{\theta}\mathbf u_0
		+
		s_{\theta}
		\dfrac{\mathbf y_{\perp}}{\|\mathbf y_{\perp}\|_2},
		&
		\begin{array}{l}
			\mathbf y\ne\mathbf 0,\ 
			\bar{\mathbf y}^{\mathrm T}\mathbf u_0<c_{\theta},\\
			\|\mathbf y_{\perp}\|_2>0,
		\end{array}
		\\[3mm]
		c_{\theta}\mathbf u_0+s_{\theta}\mathbf b_{\perp},
		&
		\text{otherwise},
	\end{cases}
	\label{eq:cap_retraction}
\end{align}
where $\mathbf b_{\perp}$ is any fixed unit vector satisfying $\mathbf b_{\perp}^{\mathrm T}\mathbf u_0=0$. For $\mathbf u_0=[0,0,1]^{\mathrm T}$, we use $\mathbf b_{\perp}=[1,0,0]^{\mathrm T}$. The mapping in \eqref{eq:cap_retraction} always returns a unit vector in $\mathcal C_{\mathrm{cap}}$. Specifically, it keeps the normalized vector unchanged when it already satisfies the spherical-cap constraint. Otherwise, it projects the vector onto the cap boundary while preserving the direction of its component orthogonal to $\mathbf u_0$.

For the $n$-th coupler, we construct two orthonormal basis vectors 
$\mathbf b_{n,1}^{(t)}$ and $\mathbf b_{n,2}^{(t)}$ for the tangent space of the unit sphere at $\mathbf u_n^{(t)}$, which satisfy
\begin{align}
	(\mathbf b_{n,r}^{(t)})^{\mathrm T}\mathbf u_n^{(t)}
	=
	0,\quad
	(\mathbf b_{n,r}^{(t)})^{\mathrm T}\mathbf b_{n,j}^{(t)}
	=
	\delta_{r,j},
	\quad r,j\in\{1,2\},
	\label{eq:tangent_basis}
\end{align}
where $\delta_{r,j}$ is the Kronecker delta. Given a small finite-difference stepsize $\epsilon_{\mathrm{fd}}>0$, the cap-feasible trial rotations along each tangent direction are generated as
\begin{subequations}
	\begin{align}
		\mathbf u_{n,r}^{+}
		&=
		\mathcal R_{\mathrm{cap}}
		\left(
		\mathbf u_n^{(t)}
		+
		\epsilon_{\mathrm{fd}}\mathbf b_{n,r}^{(t)}
		\right),
		\label{eq:un_plus}\\
		\mathbf u_{n,r}^{-}
		&=
		\mathcal R_{\mathrm{cap}}
		\left(
		\mathbf u_n^{(t)}
		-
		\epsilon_{\mathrm{fd}}\mathbf b_{n,r}^{(t)}
		\right).
		\label{eq:un_minus}
	\end{align}
\end{subequations}
Matrices $\mathbf U_{n,r}^{+}$ and $\mathbf U_{n,r}^{-}$ are obtained from $\mathbf U^{(t)}$ by replacing its $n$-th column with $\mathbf u_{n,r}^{+}$ and $\mathbf u_{n,r}^{-}$, respectively. Although each trial rotation is individually feasible with respect to the spherical-cap constraint, the resulting matrix may still violate the coupled non-intersection constraints. Therefore, the finite-difference coefficient is evaluated using feasible trial matrices as
\begin{align}
	\varsigma_{n,r}^{(t)}
	=
	\begin{cases}
		\dfrac{\Phi(\mathbf U_{n,r}^{+})-\Phi(\mathbf U_{n,r}^{-})}{2\epsilon_{\mathrm{fd}}},
		&
		\mathbf U_{n,r}^{+},\mathbf U_{n,r}^{-}\in\mathcal U,
		\\[3mm]
		\dfrac{\Phi(\mathbf U_{n,r}^{+})-\Phi(\mathbf U^{(t)})}{\epsilon_{\mathrm{fd}}},
		&
		\mathbf U_{n,r}^{+}\in\mathcal U,\ 
		\mathbf U_{n,r}^{-}\notin\mathcal U,
		\\[3mm]
		\dfrac{\Phi(\mathbf U^{(t)})-\Phi(\mathbf U_{n,r}^{-})}{\epsilon_{\mathrm{fd}}},
		&
		\mathbf U_{n,r}^{+}\notin\mathcal U,\ 
		\mathbf U_{n,r}^{-}\in\mathcal U,
		\\[3mm]
		0,
		&
		\mathbf U_{n,r}^{+},\mathbf U_{n,r}^{-}\notin\mathcal U.
	\end{cases}
	\label{eq:fd_scalar}
\end{align}
The finite-difference gradient estimate with respect to $\mathbf u_n$ is then constructed as
\begin{align}
	\widetilde{\mathbf g}_n^{(t)}
	=
	\sum_{r=1}^{2}
	\varsigma_{n,r}^{(t)}
	\mathbf b_{n,r}^{(t)}.
	\label{eq:fd_gradient}
\end{align}
This construction uses a central difference when both forward and backward trial matrices are feasible, and otherwise resorts to a one-sided feasible difference.

For numerical robustness, we project \eqref{eq:fd_gradient} onto the tangent space of the unit sphere at $\mathbf u_n^{(t)}$ as
\begin{align}
	\mathbf q_n^{(t)}
	=
	\mathbf T_n^{(t)}
	\widetilde{\mathbf g}_n^{(t)},
	\qquad
	\mathbf T_n^{(t)}
	=
	\mathbf I_3
	-
	\mathbf u_n^{(t)}(\mathbf u_n^{(t)})^{\mathrm T}.
	\label{eq:projected_gradient}
\end{align}
Here, $\mathbf T_n^{(t)}$ is the orthogonal projection matrix onto the tangent space at $\mathbf u_n^{(t)}$. The projected vector, $\mathbf q_n^{(t)}$, is then used to construct a local first-order ascent model over the spherical cap. The search point is obtained by maximizing the local first-order model directly over $\mathcal C_{\mathrm{cap}}$. Specifically, the linear oracle for the $n$-th coupler is given by 
\begin{align}
	\mathbf s_n^{(t)}
	\in
	\arg\max_{\mathbf x\in\mathcal C_{\mathrm{cap}}}
	(\mathbf q_n^{(t)})^{\mathrm T}\mathbf x,
~ n\in\mathcal N.
	\label{eq:linear_oracle_RCA}
\end{align}
The oracle in \eqref{eq:linear_oracle_RCA} has a closed-form solution. Specifically, when $\|\mathbf q_n^{(t)}\|_2=0$, the objective in \eqref{eq:linear_oracle_RCA} is constant over $\mathcal C_{\mathrm{cap}}$. In this case, the current orientation is retained, and the oracle solution is given by
\begin{align}
	\mathbf s_n^{(t)}
	=
	\mathbf u_n^{(t)}.
	\label{eq:oracle_zero_v3}
\end{align}

When $\|\mathbf q_n^{(t)}\|_2>0$, the normalized ascent direction and its perpendicular component with respect to $\mathbf u_0$ are calculated as
\begin{subequations}
	\begin{align}
		\widehat{\mathbf q}_n^{(t)}
		&=
		\frac{\mathbf q_n^{(t)}}{\|\mathbf q_n^{(t)}\|_2},
		\label{eq:qhat_v3}\\
		\mathbf q_{n,\perp}^{(t)}
		&=
		(\mathbf I_3-\mathbf u_0\mathbf u_0^{\mathrm T})
		\mathbf q_n^{(t)} .
		\label{eq:qperp_v3}
	\end{align}
\end{subequations}
If the normalized direction satisfies $(\widehat{\mathbf q}_n^{(t)})^{\mathrm T}\mathbf u_0\ge c_{\theta}$, it already lies in $\mathcal C_{\mathrm{cap}}$. Therefore, the maximizer of \eqref{eq:linear_oracle_RCA} is obtained as
\begin{align}
	\mathbf s_n^{(t)}
	=
	\widehat{\mathbf q}_n^{(t)}.
	\label{eq:oracle_inside_v3}
\end{align}

If $(\widehat{\mathbf q}_n^{(t)})^{\mathrm T}\mathbf u_0<c_{\theta}$, the maximizer lies on the boundary circle of the spherical cap. When $\|\mathbf q_{n,\perp}^{(t)}\|_2>0$, the optimal boundary point is obtained by aligning its component orthogonal to $\mathbf u_0$ with $\mathbf q_{n,\perp}^{(t)}$, namely
\begin{align}
	\mathbf s_n^{(t)}
	=
	c_{\theta}\mathbf u_0
	+
	s_{\theta}
	\frac{
		\mathbf q_{n,\perp}^{(t)}
	}{
		\|\mathbf q_{n,\perp}^{(t)}\|_2
	}.
	\label{eq:oracle_boundary_v3}
\end{align}
This solution satisfies $\|\mathbf s_n^{(t)}\|_2=1$ and $(\mathbf s_n^{(t)})^{\mathrm T}\mathbf u_0=c_{\theta}$, and thus it lies on the boundary of $\mathcal C_{\mathrm{cap}}$.

Finally, if $(\widehat{\mathbf q}_n^{(t)})^{\mathrm T}\mathbf u_0<c_{\theta}$ and $\|\mathbf q_{n,\perp}^{(t)}\|_2=0$, the boundary maximizer is not unique since the projected ascent direction has no component orthogonal to $\mathbf u_0$. In this degenerate case, the oracle solution is set as
\begin{align}
	\mathbf s_n^{(t)}
	=
	c_{\theta}\mathbf u_0
	+
	s_{\theta}\mathbf b_{\perp}.
	\label{eq:oracle_degenerate_v3}
\end{align}

After solving the oracle in \eqref{eq:linear_oracle_RCA}, the conditional-gradient direction of the $n$-th coupler is given by
\begin{align}
	\mathbf d_n^{(t)}
	=
	\mathbf s_n^{(t)}
	-
	\mathbf u_n^{(t)}.
	\label{eq:FW_direction_RCA}
\end{align}
The corresponding conditional-gradient gap is calculated as
\begin{align}
	\Delta^{(t)}
	=
	\sum_{n=1}^{N}
	(\mathbf q_n^{(t)})^{\mathrm T}
	\left(
	\mathbf s_n^{(t)}
	-
	\mathbf u_n^{(t)}
	\right).
	\label{eq:FW_gap}
\end{align}

By the optimality of $\mathbf s_n^{(t)}$ in \eqref{eq:linear_oracle_RCA}, we have
$(\mathbf q_n^{(t)})^{\mathrm T}\mathbf s_n^{(t)}
\ge
(\mathbf q_n^{(t)})^{\mathrm T}\mathbf u_n^{(t)}$
because $\mathbf u_n^{(t)}\in\mathcal C_{\mathrm{cap}}$. Hence, $\Delta^{(t)}\ge0$.
A small value of $\Delta^{(t)}$ indicates that the current rotation axis matrix is close to a stationary point under the adopted first-order model.

For a trial stepsize, $\rho\in(0,1]$, the candidate rotation of the $n$-th coupler is generated by
\begin{align}
	\widetilde{\mathbf u}_n(\rho)
	=
	\mathcal R_{\mathrm{cap}}
	\left(
	\mathbf u_n^{(t)}
	+
	\rho\mathbf d_n^{(t)}
	\right),
	\qquad n\in\mathcal N,
	\label{eq:candidate_update}
\end{align}
and the candidate rotation axis matrix is obtained as
\begin{align}
	\widetilde{\mathbf U}(\rho)
	=
	[
	\widetilde{\mathbf u}_1(\rho),
	\widetilde{\mathbf u}_2(\rho),
	\ldots,
	\widetilde{\mathbf u}_N(\rho)
	].
	\label{eq:candidate_U}
\end{align}
The retraction enforces the individual spherical-cap constraints at every trial stepsize, while the pairwise non-intersection constraints remain coupled across different couplers. Therefore, the stepsize is chosen by Armijo backtracking with an explicit feasibility check. Specifically, the Armijo parameter and the shrinkage factor are denoted by $\alpha\in(0,1)$ and $\beta\in(0,1)$, respectively. Starting from $\rho=1$, the candidate stepsize is repeatedly updated as $\rho\leftarrow\beta\rho$ until
\begin{subequations}
	\begin{align}
		\widetilde{\mathbf U}(\rho)&\in\mathcal U,
		\label{eq:armijo_feasibility}\\
		\Phi(\widetilde{\mathbf U}(\rho))
		&\ge
		\Phi(\mathbf U^{(t)})
		+
		\alpha\rho\Delta^{(t)}.
		\label{eq:armijo_sufficient_increase}
	\end{align}
\end{subequations}
The first condition in \eqref{eq:armijo_feasibility} preserves coupler physical feasibility, while the second condition in \eqref{eq:armijo_sufficient_increase} guarantees a sufficient increase in the objective value. With the accepted stepsize denoted by $\rho^{(t)}$, the rotation axis matrix is updated as
\begin{align}
	\mathbf U^{(t+1)}
	=
	\widetilde{\mathbf U}(\rho^{(t)}).
	\label{eq:U_update}
\end{align}
Thus, every accepted continuous-refinement iterate remains in $\mathcal U$ and the objective value is monotonically non-decreasing after initialization.

\subsection{Codebook-Aided Feasible Initialization}
The continuous update described above is a local refinement method and may be sensitive to the initial feasible point. To mitigate this issue, we use a discrete stochastic search over a finite spherical-cap codebook to provide a high-quality starting point for the subsequent continuous refinement.

Specifically, we construct a spherical Fibonacci codebook over $\mathcal C_{\mathrm{cap}}$, which provides nearly uniform sampling over the spherical cap~\cite{marques2013spherical}. Let $N_{\mathrm{d}}$ denote the number of codewords, and let the golden ratio be
\begin{align}
	\varphi_{\mathrm g}
	=
	\frac{1+\sqrt{5}}{2}.
	\label{eq:golden_ratio}
\end{align}
For $i=1,2,\ldots,N_{\mathrm{d}}$, the zenith angle of the $i$-th codeword is calculated as
\begin{align}
	\vartheta_i
	=
	\arccos\!\left(
	1
	-
	\frac{i-\frac{1}{2}}{N_{\mathrm{d}}}
	\left(1-\cos\theta_{\max}\right)
	\right).
	\label{eq:fib_theta}
\end{align}
The corresponding azimuth angle is given by
\begin{align}
	\varphi_i
	=
	\mathrm{mod}
	\left(
	\frac{2\pi(i-1)}{\varphi_{\mathrm g}},
	2\pi
	\right).
	\label{eq:fib_phi}
\end{align}
Then, the $i$-th codeword is expressed as
\begin{align}
	\mathbf c_i
	=
	[\sin\vartheta_i\cos\varphi_i,\,
	\sin\vartheta_i\sin\varphi_i,\,
	\cos\vartheta_i]^{\mathrm T}.
	\label{eq:ci_codeword}
\end{align}

Since $\|\mathbf c_i\|_2=1$ and $\mathbf u_0^{\mathrm T}\mathbf c_i=\cos\vartheta_i\ge\cos\theta_{\max}$, each codeword belongs to $\mathcal C_{\mathrm{cap}}$. The resulting codebook is given by
\begin{align}
	\mathcal C_{\mathrm{SF}}
	=
	\{\mathbf c_1,\mathbf c_2,\ldots,\mathbf c_{N_{\mathrm{d}}}\}.
	\label{eq:CSF_def}
\end{align}

An exhaustive search over $\mathcal C_{\mathrm{SF}}^N$ requires $N_{\mathrm{d}}^N$ objective evaluations and is impractical even for moderate $N$. We therefore adopt the CEM~\cite{rubinstein1999cross}, which iteratively learns a categorical sampling distribution that concentrates on high-quality feasible rotation combinations. At CEM iteration $r$, the probability vector for the $n$-th coupler is expressed as
\begin{align}
	\boldsymbol{\mu}_n^{(r)}
	=
	[\mu_{n,1}^{(r)},\mu_{n,2}^{(r)},\ldots,\mu_{n,N_{\mathrm{d}}}^{(r)}]^{\mathrm T},
	\label{eq:cem_pmf}
\end{align}
where $\mu_{n,i}^{(r)}$ denotes the probability of selecting codeword $\mathbf c_i$ for the $n$-th coupler at CEM iteration $r$, and satisfies
\begin{align}
	\mu_{n,i}^{(r)}\ge0,
	\qquad
	\sum_{i=1}^{N_{\mathrm{d}}}\mu_{n,i}^{(r)}=1,
	\qquad n\in\mathcal N.
	\label{eq:cem_pmf_constraint}
\end{align}
The probability distributions are initialized uniformly as
\begin{align}
	\mu_{n,i}^{(0)}
	=
	\frac{1}{N_{\mathrm{d}}},
~
	n\in\mathcal N,\quad i=1,2,\ldots,N_{\mathrm{d}}.
	\label{eq:cem_uniform}
\end{align}

During the CEM process, we maintain the feasible-sample set $\mathcal S_{\mathrm C}$ to store all feasible rotation axis matrices generated over all CEM iterations. This set is initialized as
$\mathcal S_{\mathrm C}=\emptyset$.
The CEM iterations are indexed by $r=1,2,\ldots,T_{\mathrm C}$.
At CEM iteration $r$, we generate $S_{\mathrm C}$ candidate rotation axis matrices, where $S_{\mathrm C}$ denotes the number of sampled candidates in each CEM iteration. For the $\nu$-th candidate, the codeword index of the $n$-th coupler is sampled as
\begin{align}
	\kappa_n^{(\nu,r)}
	\sim
	\mathrm{Cat}\!\left(\boldsymbol{\mu}_n^{(r-1)}\right),
	\qquad n\in\mathcal N,
	\label{eq:cem_sampling}
\end{align}
where $\mathrm{Cat}(\boldsymbol{\mu}_n^{(r-1)})$ denotes the categorical distribution with probability vector $\boldsymbol{\mu}_n^{(r-1)}$, i.e.,
$\Pr(\kappa_n^{(\nu,r)}=i)=\mu_{n,i}^{(r-1)}$ for $i=1,2,\ldots,N_{\mathrm{d}}$.
Then, the sampled rotation axis matrix is obtained as
\begin{align}
	\mathbf U^{(\nu,r)}
	=
	[
	\mathbf c_{\kappa_1^{(\nu,r)}},
	\mathbf c_{\kappa_2^{(\nu,r)}},
	\ldots,
	\mathbf c_{\kappa_N^{(\nu,r)}}
	].
	\label{eq:cem_candidate_U}
\end{align}
Although each sampled rotation axis vector $\mathbf c_{\kappa_n^{(\nu,r)}}$ belongs to $\mathcal C_{\mathrm{cap}}$, the sampled matrix $\mathbf U^{(\nu,r)}$ may still violate the coupled non-intersection constraints. Hence, the feasible sample set is defined as
\begin{align}
	\mathcal F^{(r)}
	=
	\left\{
	\nu\in\{1,2,\ldots,S_{\mathrm{C}}\}
	\,\middle|\,
	\mathbf U^{(\nu,r)}\in\mathcal U
	\right\}.
	\label{eq:CEM_feasible_index}
\end{align}
\begin{algorithm}[!t]
	\caption{Spherical-Cap Conditional-Gradient-Based Coupler Rotation Optimization Algorithm}
	\label{alg:RCA_FW}
	\begin{algorithmic}[1]
		\State \textbf{Input} $\Phi(\mathbf U)$, feasible set $\mathcal U$, fallback initialization $\mathbf U_{\mathrm{fb}}\in\mathcal U$.
		\State Construct $\mathcal C_{\mathrm{SF}}$ according to \eqref{eq:fib_theta}--\eqref{eq:CSF_def}.
		\State Obtain $\mathbf U^{(0)}$ by the CEM initialization in \eqref{eq:cem_uniform}--\eqref{eq:UCEM_best}.
		\If{$\mathcal S_{\mathrm{C}}=\emptyset$}
		\State Set $\mathbf U^{(0)}\leftarrow\mathbf U_{\mathrm{fb}}$.
		\EndIf
		\State Set $t\leftarrow0$ and compute $\Phi^{(0)}\leftarrow\Phi(\mathbf U^{(0)})$.
		\While{$t<T_{\max}$}
		\State Compute $\{\widetilde{\mathbf g}_n^{(t)}\}_{n=1}^{N}$ according to \eqref{eq:fd_scalar} and \eqref{eq:fd_gradient}.
		\State Compute $\{\mathbf q_n^{(t)}\}_{n=1}^{N}$ according to \eqref{eq:projected_gradient}.
		\State Solve the linear oracle in \eqref{eq:linear_oracle_RCA} to obtain $\{\mathbf s_n^{(t)}\}_{n=1}^{N}$.
		\State Compute $\mathbf d_n^{(t)}=\mathbf s_n^{(t)}-\mathbf u_n^{(t)}$, $\forall n\in\mathcal N$.
		\State Compute $\Delta^{(t)}$ according to \eqref{eq:FW_gap}.
		\If{$\Delta^{(t)}\le\epsilon$}
		\State \textbf{break}.
		\EndIf
		\State Find $\rho^{(t)}$ by Armijo backtracking such that \eqref{eq:armijo_feasibility} and \eqref{eq:armijo_sufficient_increase} hold.
		\If{$\rho^{(t)}<\rho_{\min}$}
		\State \textbf{break}.
		\EndIf
		\State Update $\mathbf U^{(t+1)}$ according to \eqref{eq:U_update}.
		\State Compute $\Phi^{(t+1)}\leftarrow\Phi(\mathbf U^{(t+1)})$.
		\State Set $t\leftarrow t+1$.
		\If{$\frac{|\Phi^{(t)}-\Phi^{(t-1)}|}{\max\{|\Phi^{(t-1)}|,1\}}\le\epsilon$}
		\State \textbf{break}.
		\EndIf
		\EndWhile
		\State Set $\mathbf U^{\star}\leftarrow\mathbf U^{(t)}$.
		\State \textbf{Output} Optimized rotation axis matrix $\mathbf U^{\star}$.
	\end{algorithmic}
\end{algorithm}

If $\mathcal F^{(r)}$ is empty, no feasible sample is added to the feasible-sample set $\mathcal S_{\mathrm C}$ in this iteration, and the probability vectors are left unchanged. Otherwise, the feasible samples are added to the accumulated feasible-sample set as
\begin{align}
	\mathcal S_{\mathrm{C}}
	\leftarrow
	\mathcal S_{\mathrm{C}}
	\cup
	\left\{
	\mathbf U^{(\nu,r)}
	\,\middle|\,
	\nu\in\mathcal F^{(r)}
	\right\}.
	\label{eq:CEM_accumulated_set}
\end{align}
Then, the elite feasible samples are selected according to their objective values. Let $\rho_{\mathrm e}\in(0,1]$ denote the elite fraction. The number of elite samples is given by
\begin{align}
	S_{\mathrm e}^{(r)}
	=
	\max
	\left\{
	1,
	\left\lceil
	\rho_{\mathrm e}|\mathcal F^{(r)}|
	\right\rceil
	\right\}.
	\label{eq:CEM_elite_number}
\end{align}

Let $\mathcal E^{(r)}\subseteq\mathcal F^{(r)}$ denote the index set of the $S_{\mathrm e}^{(r)}$ feasible samples with the largest objective values $\Phi(\mathbf U^{(\nu,r)})$. Consequently, for each coupler $n$ and codeword $\mathbf c_i$, the empirical elite frequency is defined as
\begin{align}
	q_{n,i}^{(r)}
	=
	\frac{
		\left|
		\left\{
		\nu\in\mathcal E^{(r)}
		\,\middle|\,
		\kappa_n^{(\nu,r)}=i
		\right\}
		\right|
	}{
		|\mathcal E^{(r)}|
	}.
	\label{eq:elite_frequency}
\end{align}
The quantity $q_{n,i}^{(r)}$ measures the fraction of elite feasible samples in which the $n$-th coupler selects codeword $\mathbf c_i$. Then, the probability vector is updated by the smoothed CEM rule as
\begin{align}
	\mu_{n,i}^{(r)}
	=
	(1-\tau_{\mathrm{C}})\mu_{n,i}^{(r-1)}
	+
	\tau_{\mathrm{C}}q_{n,i}^{(r)},
	\label{eq:cem_pmf_update}
\end{align}
for $n\in\mathcal N$ and $i=1,2,\ldots,N_{\mathrm{d}}$,
where $\tau_{\mathrm{C}}\in(0,1]$ is the smoothing factor. This update increases the probabilities of codewords that appear more frequently among elite feasible samples while avoiding abrupt probability changes.

After $T_{\mathrm C}$ iterations, the best feasible sample stored in the feasible-sample set $\mathcal S_{\mathrm C}$ is used to initialize the continuous refinement.
If $\mathcal S_{\mathrm{C}}\ne\emptyset$, the initial point for the continuous refinement is obtained as
\begin{align}
	\mathbf U^{(0)}
	=
	\arg\max_{\mathbf U\in\mathcal S_{\mathrm{C}}}
	\Phi(\mathbf U).
	\label{eq:UCEM_best}
\end{align}
If no feasible sample is found, the fixed-axis configuration $\mathbf U^{(0)}=[\mathbf u_0,\mathbf u_0,\ldots,\mathbf u_0]$ is used as a fallback initialization when feasible.
\begin{table}[!t]
	\centering
	\caption{Simulation Parameters}
	\label{tab:sim_parameters}
	\footnotesize
	\begin{tabular}{@{}lll@{}}
		\toprule
		Symbol & Description & Value\\
		\midrule
		$f_c$ & Carrier frequency & $7$ GHz\\
		$\lambda$ & Wavelength & $0.043$ m\\
		$D$ & Dipole length & $0.5\lambda$\\
		$a$ & Dipole radius & $\lambda/500$\\
		$N$ & Number of couplers & $3$\\
		$r_{\mathrm{BU}}$ & Reference transmitter-receiver distance & $250$ m\\
		$L$ & Number of channel paths & $6$\\
		$\mathbf X$ & Passive load impedance matrix & $(0.05+\mathrm j50)\mathbf I_N\Omega$\\
		$\theta_{\max}$ & Maximum rotation angle & $\pi$\\
		
		$P$ & Transmit power & $30$ dBm\\
		\bottomrule
	\end{tabular}
\end{table}
	
Algorithm~\ref{alg:RCA_FW} summarizes the proposed procedure. The CEM stage explores the discrete codebook to obtain a high-quality feasible initial point. The continuous conditional-gradient stage then refines the coupler rotations over the continuous spherical caps. Every accepted refinement iterate satisfies $\mathbf U^{(t+1)}\in\mathcal U$ because the Armijo search explicitly checks physical feasibility. Moreover, the accepted steps make $\{\Phi(\mathbf U^{(t)})\}$ nondecreasing after initialization. Since $\mathcal U$ is compact and the normalized SNR gain is bounded in practical RCA configurations, the generated objective sequence converges. 
In addition, the dominant computational cost of Algorithm~1 comes from evaluating $\Phi(\mathbf U)$, where each evaluation scales as $\mathcal O(N^3)$. Therefore, the overall computational complexity is
$
\mathcal O\left(
\left(
T_{\mathrm{C}}S_{\mathrm{C}}
+
T_{\max}(2N+S_{\mathrm A})
\right)N^3
\right)$,
where $S_{\mathrm A}$ denotes the average number of Armijo trials.

\section{Simulation Results}
In this section, numerical examples are provided to validate
the effectiveness of the proposed RCA and coupler rotation optimization algorithm. Unless otherwise specified, the simulation parameters are summarized in Table I. For comparison, the proposed RCA is evaluated against the following baseline schemes:
\begin{itemize}
	\item \textbf{Active antenna array}. 
All $N+1$ ports are active antennas and are arranged as a fixed uniform linear array (ULA) along the $x$-axis with half-wavelength inter-element spacing. They use omnidirectional responses and are modeled as thin-wire antennas parallel to the $z$-axis.
	
\item \textbf{Fixed-rotation coupler antenna}. 
This baseline uses the same fixed coupler-center locations and load impedance matrix as the proposed RCA, but fixes all couplers parallel to the $z$-axis, i.e., $\mathbf u_n=\mathbf u_0$, $\forall n\in\mathcal N$. 
	
\item \textbf{Flexible-position coupler antenna} \cite{shao2026coupler}. 
The active antenna is fixed at the origin, while the $N$ couplers translate within a square movement region of size $0.8\lambda\times0.8\lambda$ in the $x$-$y$ plane around the active antenna, with their positions optimized by the algorithm in \cite{shao2026coupler}. The active antenna and couplers use omnidirectional responses, and all elements are modeled as thin-wire dipoles parallel to the $z$-axis.
\end{itemize}
\begin{figure}[t!]
	\centering
	\setlength{\abovecaptionskip}{0.cm}
	\includegraphics[width=3.5in]{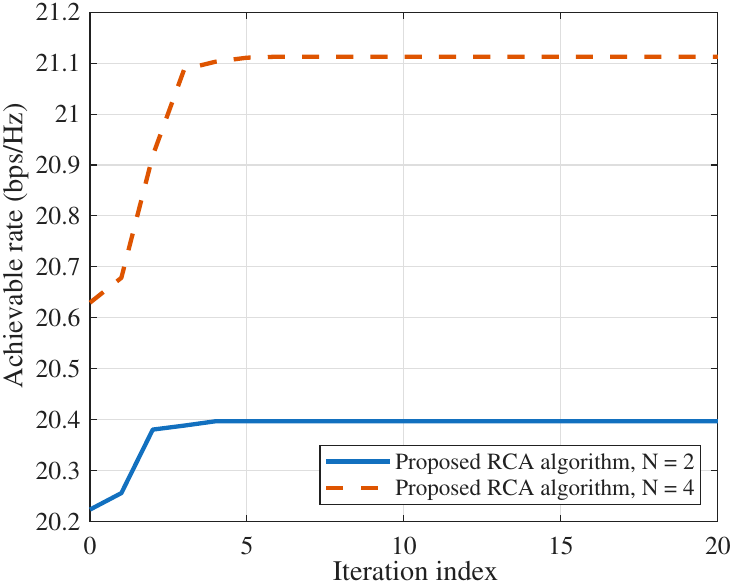}
	\caption{Convergence behavior of the proposed algorithm.}
	\label{convergence}
	\vspace{-0.59cm}
\end{figure}

Fig.~\ref{convergence} shows the convergence behavior of the proposed algorithm by plotting the achievable rate versus the iteration index. The achievable rate is given by
$R(\mathbf U)
	=
	\log_2\!\left(1+r(\mathbf U)\right)$.
The point at iteration zero corresponds to the CEM-based initialization, while the subsequent points are obtained by the continuous spherical-cap conditional-gradient refinement. It is observed that the achievable rate is improved within the first few iterations and then gradually converges, which validates the effectiveness and stable convergence behavior of the proposed algorithm. The higher rate achieved with a larger number of couplers further confirms the benefit of additional coupler rotation degrees of freedom (DoFs).
\begin{figure}[t!]
	\centering
	\setlength{\abovecaptionskip}{0.cm}
	\includegraphics[width=3.5in]{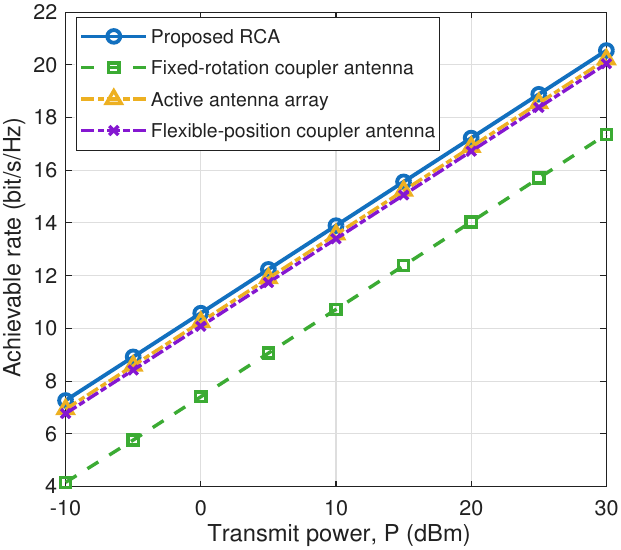}
	\caption{Achievable rate versus transmit power for different schemes.}
	\label{power}
	\vspace{-0.59cm}
\end{figure}

\begin{figure}[t!]
	\centering
	\setlength{\abovecaptionskip}{0.cm}
	\includegraphics[width=3.5in]{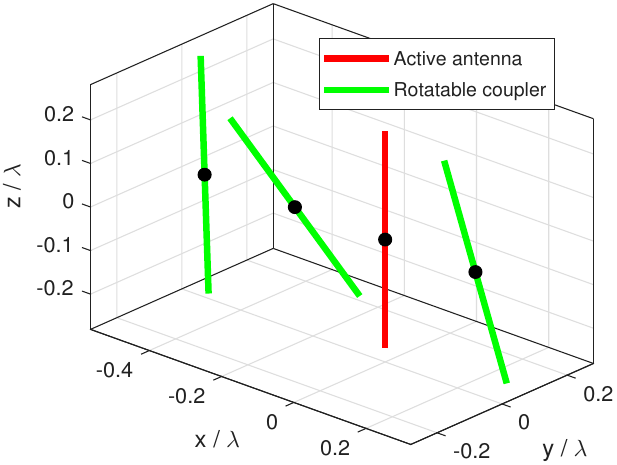}
	\caption{Optimized coupler rotations around the active antenna.}
	\label{layout}
	\vspace{-0.59cm}
\end{figure}

Fig.~\ref{power} compares the achievable rates of different schemes versus the transmit power. It is observed that, among all schemes, the proposed RCA achieves the highest achievable rate with low RF-chain cost over the whole power range. This is because rotating the couplers changes their 3D rotation axis vectors, which modifies both the orientation-dependent impedance matrix and the directional responses of the thin-wire couplers. As a result, the induced currents excited by the active antenna can be more effectively reconfigured to enhance the effective channel gain. In rich multipath environments, such reconfiguration also provides fading-mitigation gain by alleviating unfavorable signal combining and deep fading. It is also observed that the proposed RCA slightly outperforms the active antenna array and the flexible-position coupler antenna. Although the active antenna array uses $N+1$ RF chains, its array geometry and element responses are fixed, and thus it cannot exploit the directional-element gain enabled by orientation-domain reconfigurability. Moreover, the gain of the flexible-position coupler antenna relies on position reconfiguration and is limited by its omnidirectional element responses.

Fig.~\ref{layout} shows the optimized 3D coupler rotations with fixed coupler centers. The active antenna is fixed at the origin, while the  coupler centers remain fixed along the $x$-axis. Only the rotations of the couplers are optimized. It is observed that the optimized couplers rotate to different 3D orientations, demonstrating that the proposed RCA can exploit orientation-domain reconfigurability to enhance communication performance without translating any antenna element. Moreover, the optimized coupler rotations are not necessarily aligned with the user direction, since the thin-wire coupler response, mutual coupling, and transmit-power normalization jointly determine the received SNR.

\begin{figure}[t!]
	\centering
	\setlength{\abovecaptionskip}{0.cm}
	\includegraphics[width=3.5in]{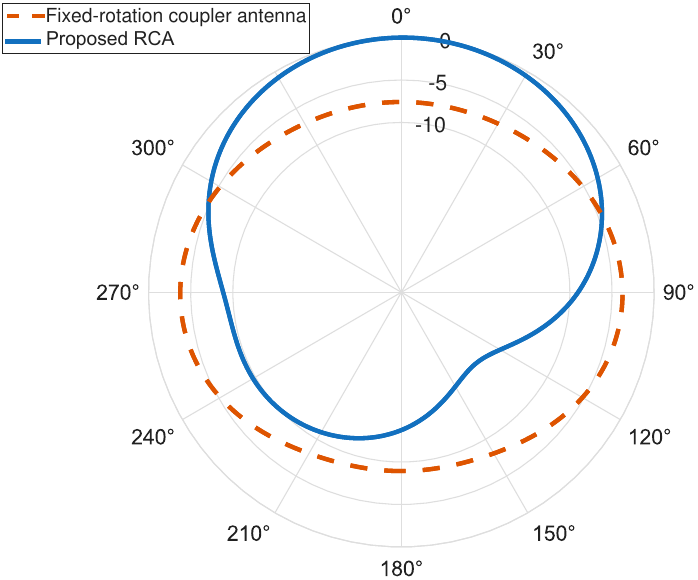}
	\caption{The normalized beampatterns (dB) of the proposed RCA and the fixed-rotation coupler antenna.}
	\label{pattern}
	\vspace{-0.59cm}
\end{figure}

Fig.~\ref{pattern} shows the normalized azimuth-plane beampatterns obtained by fixing the zenith angle as $\psi=55^\circ$ and scanning the azimuth angle $\phi$. It is observed that the proposed RCA achieves a stronger and more directional radiation pattern than the fixed-rotation coupler antenna. This is because the proposed RCA can optimize the rotations of the couplers according to the multipath channel, thereby reshaping both the orientation-dependent mutual coupling and the antenna responses. As a result, the induced currents on the couplers can be more effectively reconfigured to enhance the desired radiation. In contrast, the fixed-rotation coupler antenna keeps all couplers parallel to the active antenna, and thus cannot adapt its induced-current distribution or element responses to the propagation directions of different paths.

\begin{figure}[t!]
	\centering
	\setlength{\abovecaptionskip}{0.cm}
	\includegraphics[width=3.5in]{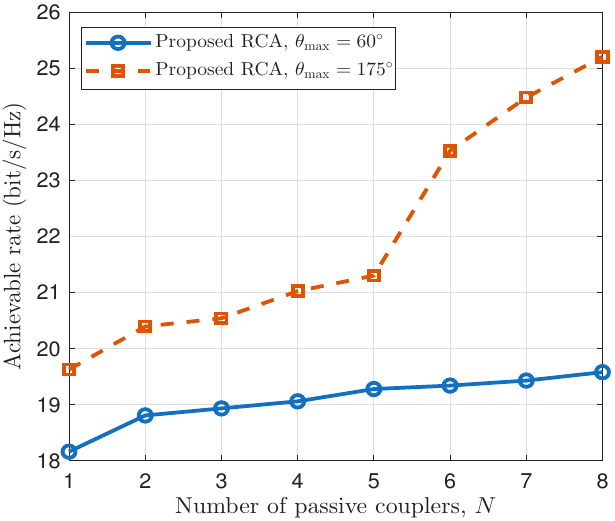}
	\caption{Achievable rates of the proposed RCA versus the number of couplers.}
	\label{N}
	\vspace{-0.46cm}
\end{figure}

Fig.~\ref{N} shows the achievable rates of the proposed RCA versus the number of couplers under different maximum rotation angles. It is observed that the achievable rate increases with $N$, since more couplers provide additional mutual-coupling interactions and orientation-domain DoFs for wireless channel reconfiguration. Moreover, the RCA with $\theta_{\max}=175^\circ$ achieves a much higher rate than that with $\theta_{\max}=60^\circ$, especially for large $N$,  because a larger rotation range enables more flexible coupler-orientation adjustment and induced-current reconfiguration.

\begin{figure}[t!]
	\centering
	\setlength{\abovecaptionskip}{0.cm}
	\includegraphics[width=3.5in]{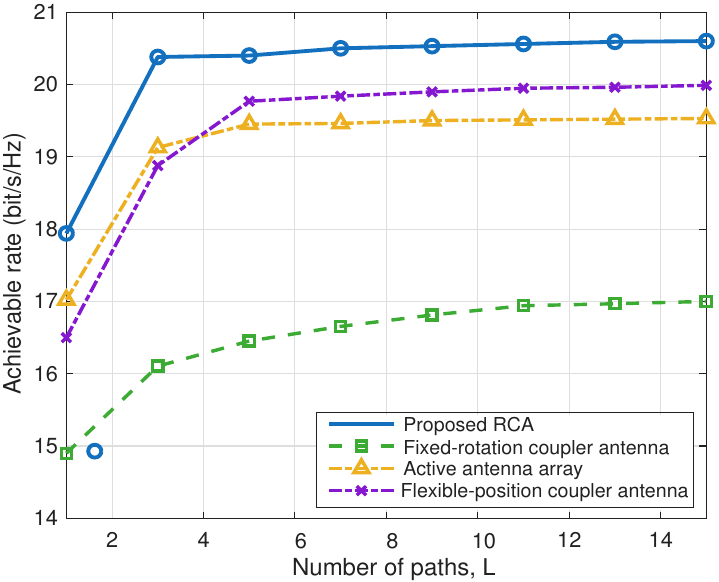}
	\caption{Achievable rates of different schemes versus the
		number of channel paths with $N=2$.}
	\label{path}
	\vspace{-0.59cm}
\end{figure}

Fig.~\ref{path} shows the achievable rates of different schemes versus the number of channel paths with $N=2$. It is observed that the achievable rates of all schemes generally increase with the number of channel paths and then gradually saturate. This is because richer multipath propagation introduces more path-dependent phase, angular, and fading variations across the transmitter aperture, which provides additional opportunities for different antenna architectures to improve the effective channel and mitigate destructive signal combining.
For the proposed RCA, coupler rotations jointly reconfigure the direction-dependent coupler responses and the induced-current distribution, thereby improving mechanical beamforming over multiple departure directions. When the number of paths becomes large, the gain gradually saturates because the available coupler-rotation DoFs are fixed and newly added correlated channel paths provide limited independent channel variations. Among all schemes, the proposed RCA consistently achieves the highest achievable rate, which verifies the effectiveness of coupler rotation in multipath propagation.

\section{Conclusions}
In this paper, we have proposed a novel RCA approach to enhance spectral and energy efficiency while reducing the implementation cost for future wireless communication systems by smartly adjusting coupler rotations. By dynamically optimizing coupler rotations via the proposed spherical-cap conditional-gradient-based algorithm with cross-entropy method initialization, the RCA can reconfigure wireless channel and directional element responses for effective mechanical beamforming. Simulation results show that the proposed RCA outperforms the fixed-geometry active antenna array while requiring only one transmit RF chain and achieves larger gains with more couplers and a wider rotation range. As a direction for further work, it is necessary to develop efficient RCA channel estimation algorithms and low-overhead coupler rotation schemes, and to extend the RCA to a six-dimensional (6D) flexible coupler antenna system with joint 3D coupler position and 3D coupler rotation.

\appendix
\section{Computation of the Mutual Impedance Matrix}
\label{app:mutual_impedance}
The entries of $\mathbf Z_{\mathrm{TX}}(\mathbf U)$ in \eqref{eq:Ztx_block} are given by
\begin{align}
	[\mathbf Z_{\mathrm{TX}}(\mathbf U)]_{m,n}
	= 
	\begin{cases}
		z_{\mathrm{s}}, & m=n,\\
		z_{m,n}(\mathbf u_m,\mathbf u_n), & m\neq n,
	\end{cases}
	~~ m,n\in\mathcal I,
	\label{eq:ZTX_entry}
\end{align}
where $z_{m,n}(\mathbf u_m,\mathbf u_n)$ denotes the mutual impedance between antenna elements $m$ and $n$.

Since the active antenna and all passive couplers have identical thin-wire geometry, their self-impedances are identical and independent of $\mathbf U$. We write the real and imaginary parts of the self-impedance as \cite{balanis2016antenna}
\begin{align}
	&\Re\{z_{\mathrm{s}}\}
	=
	\frac{\eta}{2\pi}
	\bigg[
	C+\ln(kD)-\Ci(kD)
	\nonumber\\
	&+\frac{1}{2}\sin(kD)
	\big(\Si(2kD)-2\Si(kD)\big)
	\nonumber\\
	&+\frac{1}{2}\cos(kD)
	\bigg(
	C+\ln\frac{kD}{2}
	+\Ci(2kD)-2\Ci(kD)
	\bigg)
	\bigg],
	\label{eq:zself_real}
\end{align}
and
\begin{align}
	&\Im\{z_{\mathrm{s}}\}
	=
	\frac{\eta}{4\pi}
	\bigg[
	2\Si(kD)
	+\cos(kD)
	\big(2\Si(kD)-\Si(2kD)\big)
	\nonumber\\
	&-\sin(kD)
	\bigg(
	2\Ci(kD)-\Ci(2kD)
	-\Ci\!\left(\frac{2ka^2}{D}\right)
	\bigg)
	\bigg],
	\label{eq:zself_imag}
\end{align}
where $C\simeq 0.5772$ is the Euler-Mascheroni constant, and $\Si(\cdot)$ and $\Ci(\cdot)$ denote the sine and cosine integral functions, respectively, with $\Si(x)=\int_{0}^{x}\frac{\sin t}{t}\,dt$ and $\Ci(x)=-\int_{x}^{\infty}\frac{\cos t}{t}\,dt$.

The off-diagonal entries of $\mathbf Z_{\mathrm{TX}}(\mathbf U)$ depend on the relative orientations of the corresponding antennas. Since the 3D rotations of couplers can make any two antennas non-coplanar and skewed, we compute all mutual impedances using a unified non-coplanar induced electromotive force (EMF) formulation \cite{Richmond1975Nonplanar}. Based on the antenna-axis representation in \eqref{eq:ri_s}, for two distinct antenna elements $i,j\in\mathcal I$, $i\neq j$, let $s$ and $t$ denote the signed axial parameters measured along the axes of antenna elements $i$ and $j$, respectively.
The distance between the two current-source points $\mathbf r_i(s;\mathbf u_i)$ and $\mathbf r_j(t;\mathbf u_j)$ on antenna elements $i$ and $j$, respectively, is calculated as
\begin{align}
	R_{i,j}(s,t;\mathbf u_i,\mathbf u_j)
	&=
	\left\|
	\mathbf r_i(s;\mathbf u_i)
	-\mathbf r_j(t;\mathbf u_j)
	\right\|_2
	\nonumber\\
	&=
	\left\|
	\mathbf p_i-\mathbf p_j
	+s\mathbf u_i
	-t\mathbf u_j
	\right\|_2 .
	\label{eq:Rij}
\end{align}

The normalized current distribution along each thin-wire antenna is modeled as \cite{Richmond1975Nonplanar}
\begin{align}
	I(s)
	=
	\frac{
		\sin\!\left(
		k\left(\frac{D}{2}-|s|\right)
		\right)
	}{
		\sin\!\left(\frac{kD}{2}\right)
	},
	~
	-\frac{D}{2}\le s\le \frac{D}{2}.
	\label{eq:current_distribution}
\end{align}
Accordingly, the derivative of the current distribution with respect to axial parameter $s$ is calculated as
\begin{align}
	I'(s)
	=
	-\frac{
		k\,\mathrm{sgn}(s)
		\cos\!\left(
		k\left(\frac{D}{2}-|s|\right)
		\right)
	}{
		\sin\!\left(\frac{kD}{2}\right)
	},
	\qquad s\neq 0.
	\label{eq:current_derivative}
\end{align}
Note that the value of $I'(s)$ at $s=0$ does not affect the integral and can be defined arbitrarily.

For compactness, we denote the induced-EMF integrand kernel associated with antenna elements $i$ and $j$ as
\begin{align}
	\Psi_{i,j}(s,t;\mathbf u_i,\mathbf u_j)
	\triangleq
	&\left[
	k^2 I(s)I(t)
	\mathbf u_i^{\mathrm T}\mathbf u_j
	-I'(s)I'(t)
	\right]
	\nonumber\\
	&\times
	\frac{
		e^{-\mathrm j k R_{i,j}(s,t;\mathbf u_i,\mathbf u_j)}
	}{
		R_{i,j}(s,t;\mathbf u_i,\mathbf u_j)
	}.
	\label{eq:Phi_ij}
\end{align}
Then, using the induced-EMF formulation for arbitrary non-coplanar skew thin-wire antennas~\cite{Richmond1975Nonplanar}, we compute the mutual impedance between antenna elements $i$ and $j$ as
\begin{align}
	z_{i,j}(\mathbf u_i,\mathbf u_j)
	\!=\!
	\frac{\mathrm j\eta}{4\pi k}
	\int_{-D/2}^{D/2}
	\int_{-D/2}^{D/2}
	\Psi_{i,j}(s,t;\mathbf u_i,\mathbf u_j)
	\,ds\,dt,
	i\neq j.
	\label{eq:z_non_coplanar}
\end{align}
Since \eqref{eq:z_non_coplanar} applies to any pair of antenna elements with arbitrary 3D relative orientations, it is used to compute both active-to-coupler and coupler-to-coupler mutual impedances.

\bibliographystyle{IEEEtran}
\bibliography{fabs}	
\end{document}